\documentclass[aps,prl,twocolumn,superscriptaddress,showpacs,floatfix,10pt]{revtex4-1}
\usepackage{graphicx,amssymb}
\usepackage[fleqn]{amsmath}
\usepackage{amsfonts}
\usepackage{dcolumn}% Align table columns on decimal point
\usepackage{bm}% bold math
\usepackage{braket}
\usepackage{multirow} 
\usepackage{MnSymbol}
\usepackage{color}

\begin{document}
%\begin{CJK*}{UTF8}{gbsn}

\date{\today}

\title{Can tetraneutron be a narrow resonance?}

\author{K. Fossez}
\affiliation{NSCL/FRIB Laboratory,
Michigan State University, East Lansing, Michigan 48824, USA}

\author{J. Rotureau}
\affiliation{NSCL/FRIB Laboratory,
Michigan State University, East Lansing, Michigan 48824, USA}
\affiliation{JINPA, Oak Ridge National Laboratory, Oak Ridge, TN 37831, USA}

\author{N. Michel}
\affiliation{NSCL/FRIB Laboratory,
Michigan State University, East Lansing, Michigan 48824, USA}

%\author{W. Nazarewicz}
%\affiliation{Department of Physics and Astronomy and NSCL/FRIB Laboratory,
%Michigan State University, East Lansing, Michigan 48824, USA}
%\affiliation{Institute of Theoretical Physics, Faculty of Physics, 
%University of Warsaw, Warsaw, Poland}

\author{M. P{\l}oszajczak}
\affiliation{Grand Acc\'el\'erateur National d'Ions Lourds (GANIL), CEA/DSM - CNRS/IN2P3, 
BP 55027, F-14076 Caen Cedex, France}

\newcommand{\n}[1]{{}^{#1}\text{n}}
\newcommand{\Be}[1]{{}^{#1}\text{Be}}

\newcommand{\NNLOopt}{${ \text{N2LO}_{ \text{opt} } \, }$}
\newcommand{\NNLOsat}{${ \text{N2LO}_{ \text{sat} } \, }$}

\newcommand{\vecop}[1]{\hat{ \boldsymbol{#1} }}

\begin{abstract}
	The search for a resonant four-neutron system has been revived thanks to the recent experimental hints reported in Phys. Rev. Lett. \textbf{116}, 052501 (2016) \cite{kisamori16_1463}. 
	The existence of such a system would deeply impact our understanding of nuclear matter and requires a critical investigation. 
	In this work, we study the existence of a four-neutron resonance in the quasi-stationary formalism using \textit{ab initio} techniques with various two-body chiral interactions. 
	We employ the No-Core Gamow Shell Model and the Density Matrix Renormalization Group method, both supplemented by the use of natural orbitals and a new identification technique for broad resonances. 
	We demonstrate that while the energy of the four-neutron system may be compatible with the experimental value, its width must be larger than the reported upper limit, supporting the interpretation of the experimental observation as a reaction process too short to form a nucleus.
\end{abstract}

%\pacs{21.60.Cs,	%Shell model
%	 21.10.Re,	%Collective levels
%	 21.10.Tg,	%Lifetimes, widths
%	 24.10.Eq	%Coupled-channel and distorted-wave models
%}

\maketitle
%\end{CJK*}

{
\textit{Introduction.} Whether the four-neutron system (${ \n{4} }$) exists or not is a long standing question which rests on the fact that such a system would be the result of the subtle interplay between the many-body components of the nuclear interaction, the Pauli principle and the coupling to the neutron continuum. The recent enthusiasm in the search for the ${ \n{4} }$ system started with the experimental claim \cite{marques02_1460} in 2002 that a bound ${ \n{4} }$ system could be formed in the breakup reaction: ${ {\Be{14}}^{*} \to \Be{10} + \n{4} }$. This result, though unconfirmed, stimulated theoretical investigations; most of them concluded that a bound ${ \n{4} }$ system cannot be supported by existing interactions. 
The most compelling study against the existence of a bound ${ \n{4} }$ system is given in Ref.~\cite{pieper03_1461} where a large range of modifications to the nuclear interaction (2-, 3- and 4-body components) were investigated in an \textit{ab initio} framework. 
However, the existence of a resonant ${ \n{4} }$ system has not been ruled out.

The experimental hints of a ${ \n{4} }$ resonance provided by the recent measurement \cite{kisamori16_1463} exacerbate the need for reliable nuclear interactions and \textit{ab initio} methods able to cope with couplings to the continuum. According to those experimental results, if such a state exists it would have an energy ${ E = 0.83 \pm 0.65 }$ (stat) ${ \pm }$ 1.25 (syst) MeV above the ${ \n{4} }$ threshold and a maximal width ${ \Gamma = 2.6 \, \text{MeV} }$. This large width is unlikely to correspond to a nuclear state \cite{fossez16_1335} and the question we want to address in this work is: Can a four-neutron system form a narrow resonance? In Ref.~\cite{hiyama16_1624} the possibility to form a ${ \n{4} }$ resonance by adding a phenomenological ${ T = 3/2 }$ three-body force to a realistic two-body interaction was investigated. In this study, the continuum was included and it was shown that unrealistic modifications to the nuclear interaction would be necessary to obtain a ${ \n{4} }$ system at the experimental value. Nevertheless, the importance of the Pauli principle in the ${ \n{4} }$ system, where the isospin is maximal, probably reduces the three- and four-body effects. In Ref.~\cite{shirokov16_1791}, where an \textit{ab initio} study of ${ \n{4} }$ was done in the harmonic oscillator (HO) basis and using the realistic two-body JISP16 interaction \cite{shirokov07_1753}, the energy and width of a ${ \n{4} }$ resonance were obtained from the extrapolation of phase-shifts of artificially bound ${ \n{4} }$ systems into the continuum and gave ${ E = 0.8 \, \text{MeV} }$ and ${ \Gamma = 1.4 \, \text{MeV} }$. However, couplings to the continuum at low energy below and above the particle emission thresholds may strongly affect the configuration mixing and make such extrapolations misleading for broad resonances. Results of Ref.~\cite{hiyama16_1624} that are based on the uniform complex scaling method \cite{aoyama06_1308,myo14_1047} are not affected by this issue and suggest that if a ${ \n{4} }$ system exists its energy and width should be larger than in Ref.~\cite{shirokov16_1791}. In the present study we investigate the conditions of existence of a ${ \n{4} }$ system using \textit{ab initio} methods for various two-body chiral interactions while including the continuum.
}

{
\label{sec_models}
\textit{Models and formalism.} In the present work, two different techniques are used to describe ${ \n{4} }$ in the continuum. These techniques are discussed below and both allow a consistent treatment of the couplings to the continuum by using the Berggren basis \cite{berggren68_32}. The Berggren single particle (s.p.) completeness relation includes explicitly the Gamow (resonant) states and the non-resonant scattering continuum. For each partial wave ${ c = ( \ell , j ) }$ we have:
\begin{equation}
	\sum_{i} \ket{ {u}_{c} ( {k}_{i} ) } \bra{ \tilde{u}_{c} ( {k}_{i} ) } + \int_{ \mathcal{L}_{c}^{+} } dk \, \ket{ {u}_{c} (k) } \bra{ \tilde{u}_{c} (k) } = \hat{1}_{c},
	\label{eq_Berggren_basis}
\end{equation}
where ${ \ket{ {u}_{c} ( {k}_{i} ) } }$ are the radial wave functions of resonant states and ${ \ket{ {u}_{c} (k) } }$ are the complex-energy scattering states along a contour ${ \mathcal{L}_{c}^{+} }$ in the fourth quadrant of the complex-momentum plane that surrounds the poles ${ \{ {k}_{i} \} }$ and then extends to ${ k \to +\infty }$.
In Eq.~\eqref{eq_Berggren_basis}, the tilde denotes the time-reversed states. The form of the contour is unimportant because of the Cauchy's integral theorem as long as the poles are all embedded between the real axis and the contour; details can be found in Ref.~\cite{michel09_2}.

In this work we used the No-Core Gamow Shell Model (NCGSM) which is a generalization of the No-Core Shell Model (NCSM) \cite{barrett13_688} into the complex-energy plane \cite{michel09_2,papadimitriou13_441} through the replacement of the usual HO s.p. basis by the Berggren basis. While the Hamiltonian operator is Hermitian, the Hamiltonian matrix in the NCGSM is complex-symmetric and has complex eigenvalues. An advantage of this approach based on the quasi-stationary formalism is that it fully includes the couplings to the continuum while still solving the many-body problem using configuration interaction techniques.

The second approach used in this work, the Density Matrix Renormalization Group (DMRG) method \cite{white92_488,rotureau06_15,rotureau09_140}, is an alternative way to solve the nuclear many-body problem in the continuum. In this approach, instead of building the Hamiltonian matrix in the full space and diagonalizing it as in the NCGSM, one starts from an approximate eigenstate of the Hamiltonian obtained in a small space and gradually includes the non-resonant continuum in the configuration mixing, while removing configurations with a small occupation number in the s.p. density matrix in order to keep calculations tractable at each step.

In order to further tame the computational challenge that are configuration interaction calculations in the continuum, the DMRG method has been supplemented with natural orbitals (n.o.) \cite{arxivShin16,fossez16_1793}, which are eigenstates of the s.p. density matrix associated with the targeted Hamiltonian eigenstate. A first standard DMRG run is performed, with some given truncations, to approach the final eigenstate of the Hamiltonian, then the corresponding n.o. are calculated and a new DMRG run is performed where the n.o. replace the Berggren basis states and the truncations are removed. This technique leads to an impressive gain in computational time while efficiently incorporating many-body correlations with the number of DMRG iterations. Following this development, the NCGSM has also been augmented by a similar technique where n.o. are generated from a truncated space large enough to have a decent approximation of the targeted Hamiltonian eigenstate. Then one selects the n.o. with an occupation into the density matrix ${ \eta }$ larger than some chosen value ${ 0 < { \eta }_{ \text{min} } < 1 }$ in order to replace the Berggren basis into a second run with fewer or no truncations.

In general, in both methods, the full spectrum of the Hamiltonian contains some many-body bound states and decaying resonances as well as a large number of many-body complex-energy scattering states. 
While the identification of bound states is straightforward, the extraction of resonant states from the non-resonant background is a problem that has no general solution. The invariance of physical states with respect to the definition of the non-resonant continuum can be used in small scale calculations, as for instance with the rotation angle in the uniform complex-scaling method \cite{aoyama06_1308,myo14_1047} that reveals the physical spectrum when changed. However, this approach is not practical for large configuration interaction calculations and one has to rely on a weaker criteria to extract resonances. Indeed, most resonances can be described approximately and identified unambiguously in the pole space which contains only resonant contributions (poles of the Berggren basis). Eigenstates in the full space are then identified as those having maximal overlap with pole space solutions \cite{michel02_8}. This overlap method shows its limit when considering broad decaying resonances whose energies and wave-functions are difficult to distinguish from the non-resonant continuum. To circumvent this limitation, the interaction is multiplied by a factor ${ f > 1 }$ so that the targeted state is bound and its identification immediate, and then the obtained eigenstate is used to identify the eigenstate for a smaller scaling factor ${ f' < f }$. This process is repeated until the factor equals one. In the NCGSM this is achieved by using the temporary eigenstate as a pivot in the Davidson method when diagonalizing the Hamiltonian matrix, while in the DMRG the temporary eigenvector is used to generate n.o. that are used to calculate the next eigenstate for a smaller factor ${ f' < f }$ etc. In practice this new technique preserves the unambiguous identification of resonances and extends the range of applicability of the overlap method to broader resonances.
}

{
\label{sec_results}
\textit{Results.} In the present work we considered a model space made of the ${ 0{s}_{1/2} }$ and ${ 0{p}_{3/2} }$ resonant shells (pole space), and associated non-resonant partial waves whose energies are selected along the contours in the complex-momentum plane defined by the points ${ (0,0) }$, ${ (0.15,-0.05) }$, ${ (0.3,0) }$ and ${ (4.0,0) }$ (all in ${ \text{fm}^{-1} }$), each segment being discretized by 15 points. We checked that the positions of the poles are unimportant as long as the poles are embedded by their associated complex continua. This model space is augmented by 30 non-resonant ${ {p}_{1/2} }$ partial waves along a real contour going up to ${ 4.0 \, \text{fm}^{-1} }$, and seven HO shells for the ${ d }$, ${ f }$ and ${ g }$ partial waves. The oscillator length for HO wave-functions is set at ${ b = 2.0 \, \text{fm} }$ and we verified that the effect of this parameter on our results is negligible.
The Woods-Saxon potential generating the s.p. basis was defined by the diffuseness ${ a = 0.67 \, \text{fm} }$, the radius ${ {R}_{0} = 1.9 \, \text{fm} }$, the depth ${ {V}_{0} = -27.0 \, \text{MeV} }$ and the spin-orbit term ${ {V}_{ \text{so} } = 9.5 \, \text{MeV} }$.

In the NCGSM, when allowing only two neutrons into the continuum, the predicted energies and widths of the ${ \n{4} }$ system for ${ {J}^{ \pi } = {0}^{+} }$ and for various two-body chiral interactions (N3LO \cite{entem03_1076}, \NNLOopt \cite{ekstrom13_865} and \NNLOsat \cite{ekstrom15_1766}) with different renormalization cutoffs (${ \lambda = 1.7-2.1 \, \text{fm}^{-1} }$) in ${ {V}_{\text{low k} } }$ \cite{bogner03_413,bogner03_1676} and the realistic two-body JISP16 interaction \cite{shirokov07_1753} for ${ \hbar \Omega = 20 \, \text{MeV} }$ are all consistent as shown in Tab.~\ref{tab_E_G}. 
Strictly speaking, the \NNLOsat interaction has three-body components and we only used its two-body part for a qualitative comparison. 
Moreover, the negligible influence of the renormalization cutoff of the interaction on the results indicates a weak influence of the missing induced three- and four-body forces.
\begin{table}[htb]
	%\caption{Energies and widths (in brakets) of the ${ {J}^{ \pi } = {0}^{+} }$ pole of the four-neutron system (in MeV) for various two-body chiral interactions and renormalization cutoffs, and the JISP16 interaction (${ \hbar \Omega = 20 \, \text{MeV} }$).}
	\caption{Energies and widths (in brakets) of the ${ {J}^{ \pi } = {0}^{+} }$ pole of the four-neutron system (in MeV) for various two-body interactions. The asterisk means that only the two-body part of the interaction was considered.}
	\begin{ruledtabular}
		\begin{tabular}{lccc}
			& ${ \lambda = 1.7 \, \text{fm}^{-1} }$ & ${ \lambda = 1.9 \, \text{fm}^{-1} }$ & ${ \lambda = 2.1 \, \text{fm}^{-1} }$ \\
			\hline \\[-6pt]
			N3LO     & 7.27 (3.69) & 7.28 (3.67) & 7.28 (3.69) \\
			\NNLOopt & 7.32 (3.74) & 7.33 (3.78) & 7.34 (3.95) \\
			\NNLOsat* & 7.24 (3.48) & 7.22 (3.58) & 7.27 (3.55) \\
			\hline \\[-6pt]
			JISP16   & \multicolumn{3}{c}{7.00 (3.72)}
		\end{tabular}
	\end{ruledtabular}
	\label{tab_E_G}
\end{table}

The large widths obtained with only two neutrons in the continuum (${ \Gamma \approx 3.7 \, \text{MeV} }$) already discredit the existence of the four-neutron system as a narrow resonance. However, in these calculations the width of the four-neutron system is mostly controlled by the occupation of the ${ p }$-waves. Below, we show that additional neutrons in the continuum lower the energy but do not reduce the width.

In the following, we use the N3LO two-body chiral interaction with a renormalization cutoff of ${ \lambda = 1.9 \, \text{fm}^{-1} }$. The role of the continuum in shaping the ${ \n{4} }$ system is illustrated by scaling the interaction by a factor ${ f = 2.0 }$ so that the system is artificially bound, and then one follows the evolution of the energy and width of the ${ {J}^{ \pi } = {0}^{+} }$ state when ${ f \to 1.0 }$ by step of 0.05 and for a total of 20 points as shown in Fig.~\ref{fig2}.
%%%%%%%%%%%%%%%%%%%%%%%%%%
\begin{figure}[htb]
	\includegraphics[width=1.0\linewidth]{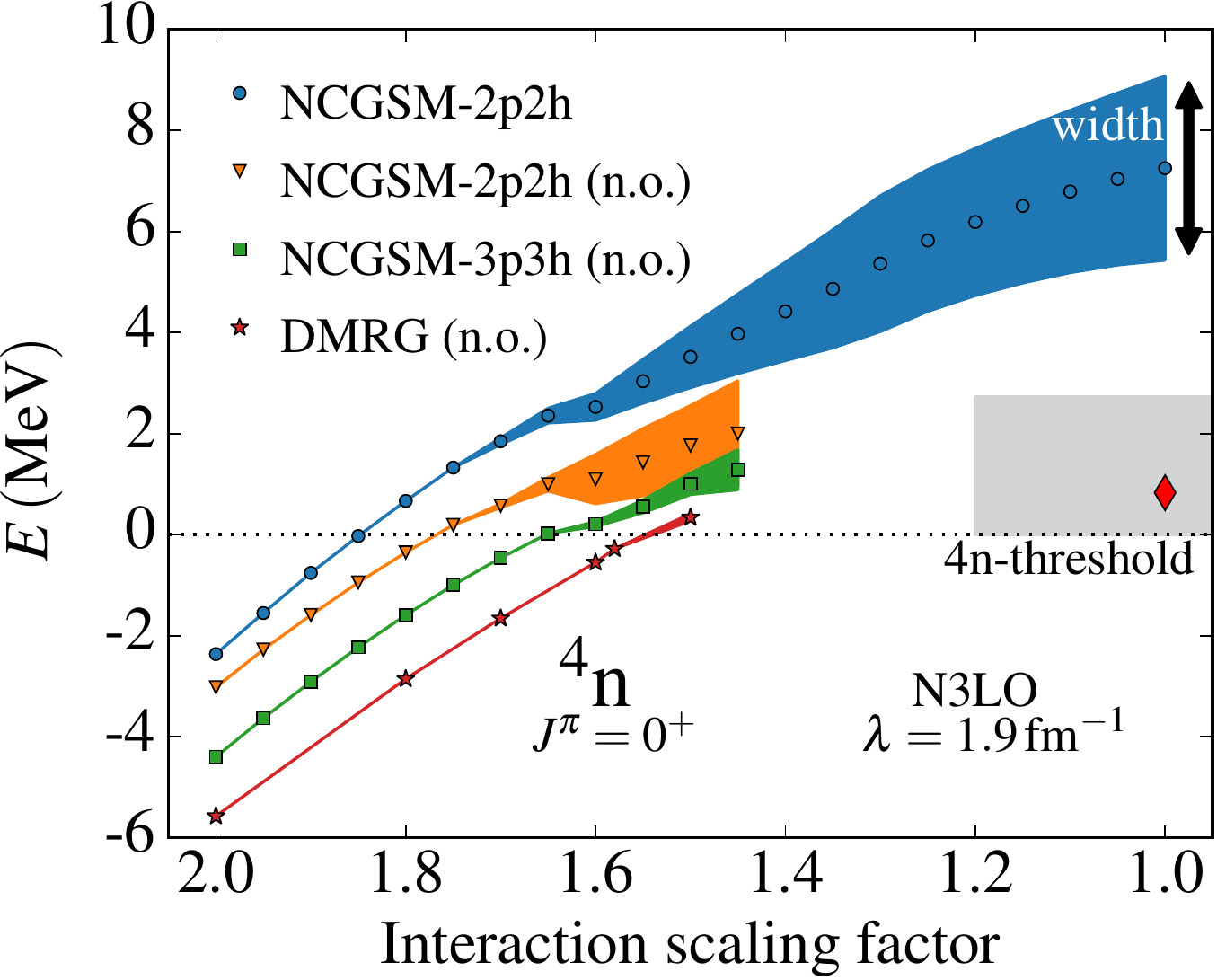}
	\caption{Evolution of the energy and width (shaded area) of the four-neutron system with the scaling of the N3LO interaction from 2.0 to 1.0. The circles represent the NCGSM results with two neutrons in the continuum, which is used to generate the NCGSM results based on natural orbitals with two (triangles) and three (squares) neutrons in the continuum. The DMRG results without truncations are represented by stars. The experimental energy is indicated by a diamond and the gray area shows the the maximal experimental uncertainties. This area is extended up to an interaction 20\% more attractive to guide the reader.}
	\label{fig2}
\end{figure}
%%%%%%%%%%%%%%%%%%%%%%%%%%

The first set of results denoted NCGSM-2p2h corresponds to the NCGSM calculations with only two neutrons in the continuum and shows a rapid increase of the width when the scaling of the interaction is gradually decreased. The small variation of the energy with the scaling factor around ${ f = 1.6 }$ is due to the use of several different basis where the ${ 0{p}_{3/2} }$ shell is bound for ${ f \geq 1.6 }$ and unbound for ${ f < 1.6 }$, to acknowledge the opening of the ${ {}^{4}\text{n} \to {}^{3}\text{n} + \text{n} }$ channel. In practice, the depth ${ {V}_{0} }$ of the potential generating the ${ p }$-waves was changed from ${ -32.0 }$ to ${ -27.0 \, \text{MeV} }$. We used the NCGSM-2p2h results to generate n.o. for each scaling factor and kept only the n.o. having an occupation ${ \eta > {10}^{-7} }$ in the density matrix, reducing the size of the basis by a factor ${ \approx 2.9 }$. The first ${ {s}_{1/2} }$ and ${ {p}_{3/2} }$ n.o. are treated as pole states while remaining orbitals are considered as continuum shells. Then we considered two and three neutrons in the n.o. continuum shells in the results denoted NCGSM-2p2h (n.o.) and NCGSM-3p3h (n.o.), respectively. This technique allows us to reduce the computational cost and to include additional many-body correlations in a more efficient way. 
The NCGSM-2p2h (n.o.) results shown in Fig.~\ref{fig2} illustrate this point clearly, as they are all lower in energy than the initial NCGSM-2p2h results obtained in the Berggren basis. However, the quality of the n.o. for the description of the targeted state depends on how close the generating eigenstate is to the final eigenstate. The NCGSM-2p2h (n.o.) calculations could only be performed for ${ f > 1.45 }$ for that reason. Another advantage of the n.o. is the possibility to remove some of the truncations as compared to the generating calculations (NCGSM-2p2h). However, the generating calculations need to include enough correlations in the continuum for the removal of truncations in the calculations with n.o. to be meaningful. In the NCGSM-3p3h (n.o.) calculations we allowed three neutrons in the n.o. outside the pole space and thus included more correlations. The NCGSM-3p3h (n.o.) calculations were limited to ${ f > 1.45 }$ as well. It was not possible to completely remove the truncations in the NCGSM and hence one had to rely on the DMRG method.

The DMRG results are without truncations on the number of particles (same shells as in the NCGSM) and the convergence criterion of the method has been fixed by the parameter ${ \varepsilon = {10}^{-8} }$ \cite{rotureau06_15,rotureau09_140}. These results are about 1 MeV lower than the NCGSM-3p3h (n.o.) at a scaling factor of ${ f = 2.0 }$, which indicates important missing correlations in the NCGSM calculations. This shows that configurations with four neutrons in the continuum shells have a large contribution to the wave-function even when the system is artificially bound. In fact, the opening of new decay channels and the presence of continuum states in the configuration mixing above the threshold is expected to make the width explode when ${ f \to 1 }$, especially in the DMRG results where all decay channels are open. This is in qualitative agreement with the results in Ref.~\cite{hiyama16_1624} which show a rapid increase of the width when the strength of the phenomenological ${ T = 3/2 }$ three-body force decreases. This explosion of the width is already visible in the NCGSM results with n.o. where, comparatively, the width increases faster than in the NCGSM-2p2h results. Another hint of the explosion of the width is the impossibility to perform the DMRG calculations far above the four-neutron threshold, even when using the improved identification technique for broad resonances. This was due to the strong couplings to the continuum, resulting in large overlaps between complex-energy scattering states and the targeted decaying resonance, making them indistinguishable. Finally, while the energy position of the four-neutron system may be compatible with the experimental value when ${ f \to 1 }$, calculations including more than two particles in the continuum as in Tab.~\ref{tab_E_G} suggest that the width of ${ \n{4} }$ is larger than ${ \Gamma \approx 3.7 \, \text{MeV} }$.
}

{
\label{sec_conclusion}
\textit{Conclusions.} In this work, we investigated the existence of the ${ \n{4} }$ system in the continuum using the No-Core Gamow Shell Model and the Density Matrix Renormalization Group method and realistic two-body interactions. Two new ingredients have been added to these approaches in order to make this study. First the introduction of natural orbitals has been a key element for improving the convergence of the calculations in the continuum, and second the progressive rescaling of the interaction to produce starting eigenstates as a technique to identify unambiguously broad resonances into the continuum was critical.

While three-body forces were not included in this work, the important role of the Pauli principle in shaping the many-body structure of the ${ \n{4} }$ system as well as its low density suggest that their exclusion yields a reasonable approximation. Interestingly, the results we obtained for various two-body chiral interactions were all consistent and were mostly dependent on the number of neutrons in the continuum. We confirm the existence of a pole of the scattering-matrix associated with the spin and parity ${ {J}^{ \pi } = {0}^{+} }$ in this system as shown in previous studies, however the proper inclusion of the couplings to the continuum shows that this pole is a feature in scattering experiments but not a genuine nuclear state. Physically this can be interpreted as a reaction process involving four neutrons which is too short to form a nucleus. However, the description of such a broad state is at the limit of the quasi-stationary formalism and it is clear that any conclusion on the existence of a light nucleus solely based on the width would be speculative to some extent. A measurement of a resonance with a half-life greater than ${ {10}^{-22} \, \text{s} }$ would provide a strong case for the existence of the ${ \n{4} }$ system as a nucleus. Finally the study of the four-neutron system pushed us to develop new techniques and opened possibilities for the study of other unbound systems such as neutron-rich hydrogen isotopes.
}

%%%%%%%%%%%%%%%%%%%%%%%%%%%%%%%%%%%%%%%%%%%%%%%%%%%%%%%%%%%%%%%%%%%%%%%%%%%%%%%%%%%%%

\begin{acknowledgments}
	We thank Witek Nazarewicz for comments and Erik Olsen for carefully reading the manuscript. We are grateful to Morten Hjorth-Jensen and James Vary for providing the \NNLOsat and JISP16 interactions, respectively. This work was supported by the U.S.\ Department of Energy, Office of Science, Office of Nuclear Physics under Awards No. DE-SC0013365 (Michigan State University) and No. DE-SC0008511 (NUCLEI SciDAC-3 collaboration), and by the National Science Foundation under award number PHY-1403906. An award of computer time was provided by the Institute for Cyber-Enabled Research at Michigan State University.
\end{acknowledgments}

%\bibliographystyle{apsrev4-1}
%%\bibliography{../../refs/apsrev_refer_new,../../refs/book}
%\bibliography{apsrev_refer}

\begin{thebibliography}{24}%
\makeatletter
\providecommand \@ifxundefined [1]{%
 \@ifx{#1\undefined}
}%
\providecommand \@ifnum [1]{%
 \ifnum #1\expandafter \@firstoftwo
 \else \expandafter \@secondoftwo
 \fi
}%
\providecommand \@ifx [1]{%
 \ifx #1\expandafter \@firstoftwo
 \else \expandafter \@secondoftwo
 \fi
}%
\providecommand \natexlab [1]{#1}%
\providecommand \enquote  [1]{``#1''}%
\providecommand \bibnamefont  [1]{#1}%
\providecommand \bibfnamefont [1]{#1}%
\providecommand \citenamefont [1]{#1}%
\providecommand \href@noop [0]{\@secondoftwo}%
\providecommand \href [0]{\begingroup \@sanitize@url \@href}%
\providecommand \@href[1]{\@@startlink{#1}\@@href}%
\providecommand \@@href[1]{\endgroup#1\@@endlink}%
\providecommand \@sanitize@url [0]{\catcode `\\12\catcode `\$12\catcode
  `\&12\catcode `\#12\catcode `\^12\catcode `\_12\catcode `\%12\relax}%
\providecommand \@@startlink[1]{}%
\providecommand \@@endlink[0]{}%
\providecommand \url  [0]{\begingroup\@sanitize@url \@url }%
\providecommand \@url [1]{\endgroup\@href {#1}{\urlprefix }}%
\providecommand \urlprefix  [0]{URL }%
\providecommand \Eprint [0]{\href }%
\providecommand \doibase [0]{http://dx.doi.org/}%
\providecommand \selectlanguage [0]{\@gobble}%
\providecommand \bibinfo  [0]{\@secondoftwo}%
\providecommand \bibfield  [0]{\@secondoftwo}%
\providecommand \translation [1]{[#1]}%
\providecommand \BibitemOpen [0]{}%
\providecommand \bibitemStop [0]{}%
\providecommand \bibitemNoStop [0]{.\EOS\space}%
\providecommand \EOS [0]{\spacefactor3000\relax}%
\providecommand \BibitemShut  [1]{\csname bibitem#1\endcsname}%
\let\auto@bib@innerbib\@empty
%</preamble>
\bibitem [{\citenamefont {Kisamori}\ \emph {et~al.}(2016)\citenamefont
  {Kisamori} \emph {et~al.}}]{kisamori16_1463}%
  \BibitemOpen
  \bibfield  {author} {\bibinfo {author} {\bibfnamefont {K.}~\bibnamefont
  {Kisamori}} \emph {et~al.},\ }\href
  {http://dx.doi.org/10.1103/PhysRevLett.116.052501} {\bibfield  {journal}
  {\bibinfo  {journal} {Phys. Rev. Lett.}\ }\textbf {\bibinfo {volume} {116}},\
  \bibinfo {pages} {052501} (\bibinfo {year} {2016})}\BibitemShut {NoStop}%
\bibitem [{\citenamefont {Marqu\'es}\ \emph {et~al.}(2002)\citenamefont
  {Marqu\'es} \emph {et~al.}}]{marques02_1460}%
  \BibitemOpen
  \bibfield  {author} {\bibinfo {author} {\bibfnamefont {F.~M.}\ \bibnamefont
  {Marqu\'es}} \emph {et~al.},\ }\href {10.1103/PhysRevC.65.044006} {\bibfield
  {journal} {\bibinfo  {journal} {Phys. Rev. C}\ }\textbf {\bibinfo {volume}
  {65}},\ \bibinfo {pages} {044006} (\bibinfo {year} {2002})}\BibitemShut
  {NoStop}%
\bibitem [{\citenamefont {Pieper}(2003)}]{pieper03_1461}%
  \BibitemOpen
  \bibfield  {author} {\bibinfo {author} {\bibfnamefont {S.~C.}\ \bibnamefont
  {Pieper}},\ }\href {https://dx.doi.org/10.1103/PhysRevLett.90.252501}
  {\bibfield  {journal} {\bibinfo  {journal} {Phys. Rev. Lett.}\ }\textbf
  {\bibinfo {volume} {90}},\ \bibinfo {pages} {252501} (\bibinfo {year}
  {2003})}\BibitemShut {NoStop}%
\bibitem [{\citenamefont {Fossez}\ \emph
  {et~al.}(2016{\natexlab{a}})\citenamefont {Fossez}, \citenamefont
  {Nazarewicz}, \citenamefont {Jaganathen}, \citenamefont {Michel},\ and\
  \citenamefont {P{\l}oszajczak}}]{fossez16_1335}%
  \BibitemOpen
  \bibfield  {author} {\bibinfo {author} {\bibfnamefont {K.}~\bibnamefont
  {Fossez}}, \bibinfo {author} {\bibfnamefont {W.}~\bibnamefont {Nazarewicz}},
  \bibinfo {author} {\bibfnamefont {Y.}~\bibnamefont {Jaganathen}}, \bibinfo
  {author} {\bibfnamefont {N.}~\bibnamefont {Michel}}, \ and\ \bibinfo {author}
  {\bibfnamefont {M.}~\bibnamefont {P{\l}oszajczak}},\ }\href
  {http://dx.doi.org/10.1103/PhysRevC.93.011305} {\bibfield  {journal}
  {\bibinfo  {journal} {Phys. Rev. C}\ }\textbf {\bibinfo {volume} {93}},\
  \bibinfo {pages} {011305(R)} (\bibinfo {year}
  {2016}{\natexlab{a}})}\BibitemShut {NoStop}%
\bibitem [{\citenamefont {Hiyama}\ \emph {et~al.}(2016)\citenamefont {Hiyama},
  \citenamefont {Lazauskas}, \citenamefont {Carbonell},\ and\ \citenamefont
  {Kamimura}}]{hiyama16_1624}%
  \BibitemOpen
  \bibfield  {author} {\bibinfo {author} {\bibfnamefont {E.}~\bibnamefont
  {Hiyama}}, \bibinfo {author} {\bibfnamefont {R.}~\bibnamefont {Lazauskas}},
  \bibinfo {author} {\bibfnamefont {J.}~\bibnamefont {Carbonell}}, \ and\
  \bibinfo {author} {\bibfnamefont {M.}~\bibnamefont {Kamimura}},\ }\href
  {http://dx.doi.org/10.1103/PhysRevC.93.044004} {\bibfield  {journal}
  {\bibinfo  {journal} {Phys. Rev. C}\ }\textbf {\bibinfo {volume} {93}},\
  \bibinfo {pages} {044004} (\bibinfo {year} {2016})}\BibitemShut {NoStop}%
\bibitem [{\citenamefont {Shirokov}\ \emph {et~al.}(2016)\citenamefont
  {Shirokov}, \citenamefont {Papadimitriou}, \citenamefont {Mazur},
  \citenamefont {Mazur}, \citenamefont {Roth},\ and\ \citenamefont
  {Vary}}]{shirokov16_1791}%
  \BibitemOpen
  \bibfield  {author} {\bibinfo {author} {\bibfnamefont {A.}~\bibnamefont
  {Shirokov}}, \bibinfo {author} {\bibfnamefont {G.}~\bibnamefont
  {Papadimitriou}}, \bibinfo {author} {\bibfnamefont {A.}~\bibnamefont
  {Mazur}}, \bibinfo {author} {\bibfnamefont {I.}~\bibnamefont {Mazur}},
  \bibinfo {author} {\bibfnamefont {R.}~\bibnamefont {Roth}}, \ and\ \bibinfo
  {author} {\bibfnamefont {J.}~\bibnamefont {Vary}},\ }\href
  {https://doi.org/10.1103/PhysRevLett.117.182502} {\bibfield  {journal}
  {\bibinfo  {journal} {Phys. Rev. Lett.}\ }\textbf {\bibinfo {volume} {117}},\
  \bibinfo {pages} {182502} (\bibinfo {year} {2016})}\BibitemShut {NoStop}%
\bibitem [{\citenamefont {Shirokov}\ \emph {et~al.}(2007)\citenamefont
  {Shirokov}, \citenamefont {Vary}, \citenamefont {Mazur},\ and\ \citenamefont
  {Weber}}]{shirokov07_1753}%
  \BibitemOpen
  \bibfield  {author} {\bibinfo {author} {\bibfnamefont {A.~M.}\ \bibnamefont
  {Shirokov}}, \bibinfo {author} {\bibfnamefont {J.~P.}\ \bibnamefont {Vary}},
  \bibinfo {author} {\bibfnamefont {A.~I.}\ \bibnamefont {Mazur}}, \ and\
  \bibinfo {author} {\bibfnamefont {T.~A.}\ \bibnamefont {Weber}},\ }\href
  {http://dx.doi.org/10.1016/j.physletb.2006.10.066} {\bibfield  {journal}
  {\bibinfo  {journal} {Phys. Lett. B}\ }\textbf {\bibinfo {volume} {644}},\
  \bibinfo {pages} {33} (\bibinfo {year} {2007})}\BibitemShut {NoStop}%
\bibitem [{\citenamefont {Aoyama}\ \emph {et~al.}(2006)\citenamefont {Aoyama},
  \citenamefont {Myo}, \citenamefont {Kat\={o}},\ and\ \citenamefont
  {Ikeda}}]{aoyama06_1308}%
  \BibitemOpen
  \bibfield  {author} {\bibinfo {author} {\bibfnamefont {S.}~\bibnamefont
  {Aoyama}}, \bibinfo {author} {\bibfnamefont {T.}~\bibnamefont {Myo}},
  \bibinfo {author} {\bibfnamefont {K.}~\bibnamefont {Kat\={o}}}, \ and\
  \bibinfo {author} {\bibfnamefont {K.}~\bibnamefont {Ikeda}},\ }\href
  {https://dx.doi.org/10.1143/PTP.116.1} {\bibfield  {journal} {\bibinfo
  {journal} {Prog. Theor. Phys.}\ }\textbf {\bibinfo {volume} {116}},\ \bibinfo
  {pages} {1} (\bibinfo {year} {2006})}\BibitemShut {NoStop}%
\bibitem [{\citenamefont {Myo}\ \emph {et~al.}(2014)\citenamefont {Myo},
  \citenamefont {Kikuchi}, \citenamefont {Masui},\ and\ \citenamefont
  {Kat\={o}}}]{myo14_1047}%
  \BibitemOpen
  \bibfield  {author} {\bibinfo {author} {\bibfnamefont {T.}~\bibnamefont
  {Myo}}, \bibinfo {author} {\bibfnamefont {Y.}~\bibnamefont {Kikuchi}},
  \bibinfo {author} {\bibfnamefont {H.}~\bibnamefont {Masui}}, \ and\ \bibinfo
  {author} {\bibfnamefont {K.}~\bibnamefont {Kat\={o}}},\ }\href
  {https://dx.doi.org/10.1016/j.ppnp.2014.08.001} {\bibfield  {journal}
  {\bibinfo  {journal} {Prog. Part. Nucl. Phys.}\ }\textbf {\bibinfo {volume}
  {79}},\ \bibinfo {pages} {1} (\bibinfo {year} {2014})}\BibitemShut {NoStop}%
\bibitem [{\citenamefont {Berggren}(1968)}]{berggren68_32}%
  \BibitemOpen
  \bibfield  {author} {\bibinfo {author} {\bibfnamefont {T.}~\bibnamefont
  {Berggren}},\ }\href {https://dx.doi.org/10.1016/0375-9474(68)90593-9}
  {\bibfield  {journal} {\bibinfo  {journal} {Nucl. Phys. A}\ }\textbf
  {\bibinfo {volume} {109}},\ \bibinfo {pages} {265} (\bibinfo {year}
  {1968})}\BibitemShut {NoStop}%
\bibitem [{\citenamefont {Michel}\ \emph {et~al.}(2009)\citenamefont {Michel},
  \citenamefont {Nazarewicz}, \citenamefont {P{\l}oszajczak},\ and\
  \citenamefont {Vertse}}]{michel09_2}%
  \BibitemOpen
  \bibfield  {author} {\bibinfo {author} {\bibfnamefont {N.}~\bibnamefont
  {Michel}}, \bibinfo {author} {\bibfnamefont {W.}~\bibnamefont {Nazarewicz}},
  \bibinfo {author} {\bibfnamefont {M.}~\bibnamefont {P{\l}oszajczak}}, \ and\
  \bibinfo {author} {\bibfnamefont {T.}~\bibnamefont {Vertse}},\ }\href
  {https://dx.doi.org/10.1088/0954-3899/36/1/013101} {\bibfield  {journal}
  {\bibinfo  {journal} {J. Phys. G}\ }\textbf {\bibinfo {volume} {36}},\
  \bibinfo {pages} {013101} (\bibinfo {year} {2009})}\BibitemShut {NoStop}%
\bibitem [{\citenamefont {Barrett}\ \emph {et~al.}(2013)\citenamefont
  {Barrett}, \citenamefont {Navr\'atil},\ and\ \citenamefont
  {Vary}}]{barrett13_688}%
  \BibitemOpen
  \bibfield  {author} {\bibinfo {author} {\bibfnamefont {B.~R.}\ \bibnamefont
  {Barrett}}, \bibinfo {author} {\bibfnamefont {P.}~\bibnamefont {Navr\'atil}},
  \ and\ \bibinfo {author} {\bibfnamefont {J.~P.}\ \bibnamefont {Vary}},\
  }\href {https://dx.doi.org/10.1016/j.ppnp.2012.10.003} {\bibfield  {journal}
  {\bibinfo  {journal} {Prog. Part. Nucl. Phys.}\ }\textbf {\bibinfo {volume}
  {69}},\ \bibinfo {pages} {131} (\bibinfo {year} {2013})}\BibitemShut
  {NoStop}%
\bibitem [{\citenamefont {Papadimitriou}\ \emph {et~al.}(2013)\citenamefont
  {Papadimitriou}, \citenamefont {Rotureau}, \citenamefont {Michel},
  \citenamefont {P{\l}oszajczak},\ and\ \citenamefont
  {Barrett}}]{papadimitriou13_441}%
  \BibitemOpen
  \bibfield  {author} {\bibinfo {author} {\bibfnamefont {G.}~\bibnamefont
  {Papadimitriou}}, \bibinfo {author} {\bibfnamefont {J.}~\bibnamefont
  {Rotureau}}, \bibinfo {author} {\bibfnamefont {N.}~\bibnamefont {Michel}},
  \bibinfo {author} {\bibfnamefont {M.}~\bibnamefont {P{\l}oszajczak}}, \ and\
  \bibinfo {author} {\bibfnamefont {B.~R.}\ \bibnamefont {Barrett}},\ }\href
  {https://dx.doi.org/10.1103/PhysRevC.88.044318} {\bibfield  {journal}
  {\bibinfo  {journal} {Phys. Rev. C}\ }\textbf {\bibinfo {volume} {88}},\
  \bibinfo {pages} {044318} (\bibinfo {year} {2013})}\BibitemShut {NoStop}%
\bibitem [{\citenamefont {White}(1992)}]{white92_488}%
  \BibitemOpen
  \bibfield  {author} {\bibinfo {author} {\bibfnamefont {S.~R.}\ \bibnamefont
  {White}},\ }\href {https://dx.doi.org/10.1103/PhysRevLett.69.2863} {\bibfield
   {journal} {\bibinfo  {journal} {Phys. Rev. Lett.}\ }\textbf {\bibinfo
  {volume} {69}},\ \bibinfo {pages} {2863} (\bibinfo {year}
  {1992})}\BibitemShut {NoStop}%
\bibitem [{\citenamefont {Rotureau}\ \emph {et~al.}(2006)\citenamefont
  {Rotureau}, \citenamefont {Michel}, \citenamefont {Nazarewicz}, \citenamefont
  {P{\l}oszajczak},\ and\ \citenamefont {Dukelsky}}]{rotureau06_15}%
  \BibitemOpen
  \bibfield  {author} {\bibinfo {author} {\bibfnamefont {J.}~\bibnamefont
  {Rotureau}}, \bibinfo {author} {\bibfnamefont {N.}~\bibnamefont {Michel}},
  \bibinfo {author} {\bibfnamefont {W.}~\bibnamefont {Nazarewicz}}, \bibinfo
  {author} {\bibfnamefont {M.}~\bibnamefont {P{\l}oszajczak}}, \ and\ \bibinfo
  {author} {\bibfnamefont {J.}~\bibnamefont {Dukelsky}},\ }\href
  {https://dx.doi.org/10.1103/PhysRevLett.97.110603} {\bibfield  {journal}
  {\bibinfo  {journal} {Phys. Rev. Lett.}\ }\textbf {\bibinfo {volume} {97}},\
  \bibinfo {pages} {110603} (\bibinfo {year} {2006})}\BibitemShut {NoStop}%
\bibitem [{\citenamefont {Rotureau}\ \emph {et~al.}(2009)\citenamefont
  {Rotureau}, \citenamefont {Michel}, \citenamefont {Nazarewicz}, \citenamefont
  {P{\l}oszajczak},\ and\ \citenamefont {Dukelsky}}]{rotureau09_140}%
  \BibitemOpen
  \bibfield  {author} {\bibinfo {author} {\bibfnamefont {J.}~\bibnamefont
  {Rotureau}}, \bibinfo {author} {\bibfnamefont {N.}~\bibnamefont {Michel}},
  \bibinfo {author} {\bibfnamefont {W.}~\bibnamefont {Nazarewicz}}, \bibinfo
  {author} {\bibfnamefont {M.}~\bibnamefont {P{\l}oszajczak}}, \ and\ \bibinfo
  {author} {\bibfnamefont {J.}~\bibnamefont {Dukelsky}},\ }\href
  {https://dx.doi.org/10.1103/PhysRevC.79.014304} {\bibfield  {journal}
  {\bibinfo  {journal} {Phys. Rev. C}\ }\textbf {\bibinfo {volume} {79}},\
  \bibinfo {pages} {014304} (\bibinfo {year} {2009})}\BibitemShut {NoStop}%
\bibitem [{\citenamefont {Shin}\ \emph {et~al.}(2016)\citenamefont {Shin},
  \citenamefont {Kim}, \citenamefont {Maris}, \citenamefont {Vary},
  \citenamefont {Forss\'en}, \citenamefont {Rotureau},\ and\ \citenamefont
  {Michel}}]{arxivShin16}%
  \BibitemOpen
  \bibfield  {author} {\bibinfo {author} {\bibfnamefont {I.~J.}\ \bibnamefont
  {Shin}}, \bibinfo {author} {\bibfnamefont {Y.}~\bibnamefont {Kim}}, \bibinfo
  {author} {\bibfnamefont {P.}~\bibnamefont {Maris}}, \bibinfo {author}
  {\bibfnamefont {J.~P.}\ \bibnamefont {Vary}}, \bibinfo {author}
  {\bibfnamefont {C.}~\bibnamefont {Forss\'en}}, \bibinfo {author}
  {\bibfnamefont {J.}~\bibnamefont {Rotureau}}, \ and\ \bibinfo {author}
  {\bibfnamefont {N.}~\bibnamefont {Michel}},\ }\href@noop {} {}\bibinfo
  {howpublished} {\url{https://arxiv.org/abs/1605.02819v1}} (\bibinfo {year}
  {2016})\BibitemShut {NoStop}%
\bibitem [{\citenamefont {Fossez}\ \emph
  {et~al.}(2016{\natexlab{b}})\citenamefont {Fossez}, \citenamefont {Rotureau},
  \citenamefont {Michel}, \citenamefont {Liu},\ and\ \citenamefont
  {Nazarewicz}}]{fossez16_1793}%
  \BibitemOpen
  \bibfield  {author} {\bibinfo {author} {\bibfnamefont {K.}~\bibnamefont
  {Fossez}}, \bibinfo {author} {\bibfnamefont {J.}~\bibnamefont {Rotureau}},
  \bibinfo {author} {\bibfnamefont {N.}~\bibnamefont {Michel}}, \bibinfo
  {author} {\bibfnamefont {Q.}~\bibnamefont {Liu}}, \ and\ \bibinfo {author}
  {\bibfnamefont {W.}~\bibnamefont {Nazarewicz}},\ }\href
  {https://doi.org/10.1103/PhysRevC.94.054302} {\bibfield  {journal} {\bibinfo
  {journal} {Phys. Rev. C}\ }\textbf {\bibinfo {volume} {94}},\ \bibinfo
  {pages} {054302} (\bibinfo {year} {2016}{\natexlab{b}})}\BibitemShut
  {NoStop}%
\bibitem [{\citenamefont {Michel}\ \emph {et~al.}(2002)\citenamefont {Michel},
  \citenamefont {Nazarewicz}, \citenamefont {P{\l}oszajczak},\ and\
  \citenamefont {Bennaceur}}]{michel02_8}%
  \BibitemOpen
  \bibfield  {author} {\bibinfo {author} {\bibfnamefont {N.}~\bibnamefont
  {Michel}}, \bibinfo {author} {\bibfnamefont {W.}~\bibnamefont {Nazarewicz}},
  \bibinfo {author} {\bibfnamefont {M.}~\bibnamefont {P{\l}oszajczak}}, \ and\
  \bibinfo {author} {\bibfnamefont {K.}~\bibnamefont {Bennaceur}},\ }\href
  {https://dx.doi.org/10.1103/PhysRevLett.89.042502} {\bibfield  {journal}
  {\bibinfo  {journal} {Phys. Rev. Lett.}\ }\textbf {\bibinfo {volume} {89}},\
  \bibinfo {pages} {042502} (\bibinfo {year} {2002})}\BibitemShut {NoStop}%
\bibitem [{\citenamefont {Entem}\ and\ \citenamefont
  {Machleidt}(2003)}]{entem03_1076}%
  \BibitemOpen
  \bibfield  {author} {\bibinfo {author} {\bibfnamefont {D.~R.}\ \bibnamefont
  {Entem}}\ and\ \bibinfo {author} {\bibfnamefont {R.}~\bibnamefont
  {Machleidt}},\ }\href {https://dx.doi.org/10.1103/PhysRevC.68.041001}
  {\bibfield  {journal} {\bibinfo  {journal} {Phys. Rev. C}\ }\textbf {\bibinfo
  {volume} {68}},\ \bibinfo {pages} {041001(R)} (\bibinfo {year}
  {2003})}\BibitemShut {NoStop}%
\bibitem [{\citenamefont {Ekstr\"om}\ \emph {et~al.}(2013)\citenamefont
  {Ekstr\"om}, \citenamefont {Baardsen}, \citenamefont {Forss\'en},
  \citenamefont {Hagen}, \citenamefont {{Hjorth-Jensen}}, \citenamefont
  {Jansen}, \citenamefont {Machleidt}, \citenamefont {Nazarewicz},
  \citenamefont {Papenbrock}, \citenamefont {Sarich},\ and\ \citenamefont
  {Wild}}]{ekstrom13_865}%
  \BibitemOpen
  \bibfield  {author} {\bibinfo {author} {\bibfnamefont {A.}~\bibnamefont
  {Ekstr\"om}}, \bibinfo {author} {\bibfnamefont {G.}~\bibnamefont {Baardsen}},
  \bibinfo {author} {\bibfnamefont {C.}~\bibnamefont {Forss\'en}}, \bibinfo
  {author} {\bibfnamefont {G.}~\bibnamefont {Hagen}}, \bibinfo {author}
  {\bibfnamefont {M.}~\bibnamefont {{Hjorth-Jensen}}}, \bibinfo {author}
  {\bibfnamefont {G.~R.}\ \bibnamefont {Jansen}}, \bibinfo {author}
  {\bibfnamefont {R.}~\bibnamefont {Machleidt}}, \bibinfo {author}
  {\bibfnamefont {W.}~\bibnamefont {Nazarewicz}}, \bibinfo {author}
  {\bibfnamefont {T.}~\bibnamefont {Papenbrock}}, \bibinfo {author}
  {\bibfnamefont {J.}~\bibnamefont {Sarich}}, \ and\ \bibinfo {author}
  {\bibfnamefont {S.~M.}\ \bibnamefont {Wild}},\ }\href
  {https://dx.doi.org/10.1103/PhysRevLett.110.192502} {\bibfield  {journal}
  {\bibinfo  {journal} {Phys. Rev. Lett.}\ }\textbf {\bibinfo {volume} {110}},\
  \bibinfo {pages} {192502} (\bibinfo {year} {2013})}\BibitemShut {NoStop}%
\bibitem [{\citenamefont {Ekstr\"om}\ \emph {et~al.}(2015)\citenamefont
  {Ekstr\"om}, \citenamefont {Jansen}, \citenamefont {Wendt}, \citenamefont
  {Hagen}, \citenamefont {Papenbrock}, \citenamefont {Carlsson}, \citenamefont
  {Forss\'en}, \citenamefont {{Hjorth-Jensen}}, \citenamefont {Navr\'atil},\
  and\ \citenamefont {Nazarewicz}}]{ekstrom15_1766}%
  \BibitemOpen
  \bibfield  {author} {\bibinfo {author} {\bibfnamefont {A.}~\bibnamefont
  {Ekstr\"om}}, \bibinfo {author} {\bibfnamefont {G.~R.}\ \bibnamefont
  {Jansen}}, \bibinfo {author} {\bibfnamefont {K.~A.}\ \bibnamefont {Wendt}},
  \bibinfo {author} {\bibfnamefont {G.}~\bibnamefont {Hagen}}, \bibinfo
  {author} {\bibfnamefont {T.}~\bibnamefont {Papenbrock}}, \bibinfo {author}
  {\bibfnamefont {B.~D.}\ \bibnamefont {Carlsson}}, \bibinfo {author}
  {\bibfnamefont {C.}~\bibnamefont {Forss\'en}}, \bibinfo {author}
  {\bibfnamefont {M.}~\bibnamefont {{Hjorth-Jensen}}}, \bibinfo {author}
  {\bibfnamefont {P.}~\bibnamefont {Navr\'atil}}, \ and\ \bibinfo {author}
  {\bibfnamefont {W.}~\bibnamefont {Nazarewicz}},\ }\href
  {http://dx.doi.org/10.1103/PhysRevC.91.051301} {\bibfield  {journal}
  {\bibinfo  {journal} {Phys. Rev. C}\ }\textbf {\bibinfo {volume} {91}},\
  \bibinfo {pages} {051301(R)} (\bibinfo {year} {2015})}\BibitemShut {NoStop}%
\bibitem [{\citenamefont {Bogner}\ \emph
  {et~al.}(2003{\natexlab{a}})\citenamefont {Bogner}, \citenamefont {Kuo},\
  and\ \citenamefont {Schwenk}}]{bogner03_413}%
  \BibitemOpen
  \bibfield  {author} {\bibinfo {author} {\bibfnamefont {S.~K.}\ \bibnamefont
  {Bogner}}, \bibinfo {author} {\bibfnamefont {T.~T.~S.}\ \bibnamefont {Kuo}},
  \ and\ \bibinfo {author} {\bibfnamefont {A.}~\bibnamefont {Schwenk}},\ }\href
  {https://dx.doi.org/10.1016/j.physrep.2003.07.001} {\bibfield  {journal}
  {\bibinfo  {journal} {Phys. Rep.}\ }\textbf {\bibinfo {volume} {386}},\
  \bibinfo {pages} {1} (\bibinfo {year} {2003}{\natexlab{a}})}\BibitemShut
  {NoStop}%
\bibitem [{\citenamefont {Bogner}\ \emph
  {et~al.}(2003{\natexlab{b}})\citenamefont {Bogner}, \citenamefont {Kuo},
  \citenamefont {Schwenk}, \citenamefont {Entem},\ and\ \citenamefont
  {Machleidt}}]{bogner03_1676}%
  \BibitemOpen
  \bibfield  {author} {\bibinfo {author} {\bibfnamefont {S.~K.}\ \bibnamefont
  {Bogner}}, \bibinfo {author} {\bibfnamefont {T.~T.~S.}\ \bibnamefont {Kuo}},
  \bibinfo {author} {\bibfnamefont {A.}~\bibnamefont {Schwenk}}, \bibinfo
  {author} {\bibfnamefont {D.~R.}\ \bibnamefont {Entem}}, \ and\ \bibinfo
  {author} {\bibfnamefont {R.}~\bibnamefont {Machleidt}},\ }\href
  {https://dx.doi.org/10.1016/j.physletb.2003.10.012} {\bibfield  {journal}
  {\bibinfo  {journal} {Phys. Lett. B}\ }\textbf {\bibinfo {volume} {576}},\
  \bibinfo {pages} {265} (\bibinfo {year} {2003}{\natexlab{b}})}\BibitemShut
  {NoStop}%
\end{thebibliography}


\end{document}